\documentclass{aa}
\usepackage{graphicx}
\begin{document}
   \title{VLT/ISAAC H-band spectroscopy of embedded massive YSOs}


   \author{T.R. Kendall
          \inst{1}
          \and W.J. de Wit
          \inst{2}
          \and J.L. Yun
          \inst{1}
          }

   \offprints{T.R. Kendall}

   \institute{Centro de Astronomia e Astrof\'{i}sica da Universidade de Lisboa,
              Observat\'{o}rio Astron\'{o}mico de Lisboa, Tapada da Ajuda,
              1349-018 Lisboa, Portugal \\ \email{tkendall@oal.ul.pt} \and
              Osservatorio Astrofisico di Arcetri, Largo E. Fermi 5, 50125
              Florence, Italy\\ \email{dewit@arcetri.astro.it} }

   \date{Received; accepted}

   \abstract{We have performed intermediate resolution (R$\sim$5000), high
signal-to-noise H-band spectroscopy of a small, initial sample of three massive
embedded young stellar objects (YSOs), using VLT/ISAAC. The sample has been
selected from sources characterised in previous literature as being likely of OB
spectral type, to be unambiguously associated with bright (H$\leq$14) single
point sources in the 2MASS database, and to have no optical counterparts. Of the
targets observed, one object shows a $\sim$B3 spectrum, similar to a main
sequence object of the same spectral type. A second object exhibits weak
He\,{\sc i} and H emission, indicating an early-type source: we detect He\,{\sc
ii} absorption, which supports a previous indirect derivation of the spectral
type as mid-O. The third object does not show absorption lines,
so no spectral type can de derived. It does, however, exhibit a rich spectrum of
strong, broad emission lines and is likely to be surrounded by dense
circumstellar material and at a very early evolutionary stage.  Our results from
this very small sample are in agreement with those of \cite{kap02}, who also
find spectra similar to optically visible main sequence stars, together with
emission line objects representing a very early evolutionary phase, in their
much larger sample of K-band spectra.   }

\titlerunning{H-band spectroscopy of YSOs}

\maketitle


\section{Introduction}

Little is known about the earliest stages of the formation of massive stars.  It
is unclear whether the general picture applicable to the formation of low-mass
stars, i.e. an accretion disk and infall onto the central star (\cite{Shu87})
can be scaled up to masses higher than $\sim$10M\sun\, (in the case of spherically symmetric accretion), as radiation pressure from
the high mass stellar core on dust mixed with the infalling gas prevents further
accretion.
Alternative explanations involving the coalescence of intermediate
mass protostars in the dense central regions of rich clusters have been proposed
(\cite{bon98}).

Observationally, the earliest stages of massive star formation are only
accessible by infrared and radio techniques, because of the presence of $\sim$10
-- 100 magnitudes of visual extinction. Ultracompact (UC) H\,{\sc ii} regions
represent this earliest phase in the evolution of a very massive star.  UC
H\,{\sc ii} regions are typically identified using radio surveys
(e.g. \cite{bro96}; \cite{wal99}).  Massive YSOs capable of ionizing a UC H\,{\sc ii}
region are recognisable by their extremely red IRAS colours
(\cite{woo89}; \cite{ost97}), and the IRAS database has become a rich source to be mined
for intermediate and high-mass YSOs using colour selection criteria (e.g. 
\cite{per87}; \cite{woo89}; \cite{cam89}; \cite{cha96}).

Despite this, measurements of the fundamental parameters of UC H\,{\sc ii}
regions, and more importantly those of the massive embedded stars which ionize
the region, cannot easily be derived from the radio and far-IR observations.
This is due principally to uncertainties in distances and in the indirect
methods used to convert radio continuum or IRAS fluxes to the physical
properties of the embedded, ionizing OB star (Walsh et al. 1997; \cite{kap02} and references therein).

To circumvent these difficulties, observational efforts are now beginning to
probe directly the photospheres of massive embedded YSOs using observations in
the near-infrared. Such wavelengths are accessible from the ground (1 --
5\,$\mu$m, JHKLM bands) and the extinction owing to the gas and dust in which the
YSO is embedded is much less than in the optical. Moreover, at thermal infrared
wavelengths, emission from the dust becomes too great, and the flux from the
photosphere itself can no longer be detected. The first spectral classification
of the ionizing star of a UC H\,{\sc ii} region was made by \cite{wat97} in the
K-band.

While an important aim of the present paper is to clarify the status
of the three objects discussed here as massive embedded YSOs, a second no less important aim is to show that intermediate resolution spectroscopy in the H-band (where veiling from circumstellar emission is minimal) is a necessary and valuable tool to detect, classify and study the photospheres of such objects. This paper attempts to build on recent work on the classification of OB stars in the H-band, and to apply it to highly obscured objects.

\subsection{Classification spectroscopy of OB stars in the H-band}

Recent developments in classification spectroscopy of OB stars in the H- and
K-bands suggest application to objects too deeply obscured to be optically
visible. In the earliest work, both \cite{lan92} and \cite{dal96} presented
libraries of stellar spectra in the H-band, but these were largely directed to
stellar population analyses. More recently, the most important study directly
relevant here has been that of \cite{han98}, who have identified a number of
H-band temperature and luminosity diagnostics for OB stars in R=2000 spectra,
involving the ratio of Br 11 (1.681\,$\mu$m) to He\,{\sc i} (1.700\,$\mu$m) and
He\,{\sc ii} (1.693\,$\mu$m). Using a full grid of spectra representing both dwarf
and supergiant luminosities, these authors find that the presence of all three
lines definitively indicates a spectral type earlier than late O. With
combinations of only two lines present, in various strengths, dwarfs can be
divided into O9 -- B1, B2 -- B5, and late B -- early A type. 

This work overlaps with that of \cite{mey98}, who have observed 85 MK standards
of near-solar metallicity, at R$\sim$3000 in the H-band. They find a number of
strong T$_{\rm eff}$ and luminosity-sensitive lines in the H-band spectra of
stars of spectral type A, and later, all the way down to M5.
These two studies show that the classification of heavily embedded YSOs, and the
near-IR study of their photospheres, could be carried out in the H-band for all
spectral types from O to M, using intermediate resolution, high signal-to-noise
spectra. This could be advantageous, since the veiling of photospheric
absorption lines by continuum emission from hot circumstellar dust is much less
severe than in the K-band (Meyer et al. 1998; \cite{ish01}). Taken together,
these studies suggest that while
K-band spectroscopy has been extremely successful in studies of low-mass YSOs
(see \cite{gre96} for one example among many) classification in the range B0 --
G0 is best performed in the H-band, as pointed out by \cite{luh98}.

\begin{table}
 \begin{flushleft}
 \caption[]{Basic and derived data for observed targets. Magnitudes and colours
 are taken from the 2MASS database, except for IRAS\,17175-3544 by Tapia et al. (1996).}
 \begin{tabular}{|l|l|l|l|l|l|l|}\hline
 IRAS & H & {\it J-H} & {\it H-K} & A$_{\rm v}$ & Sp.T. & Int./s \\ \hline
 17175-3544$^a$ & 14.00 & 2.75 & 3.43 & $>$13$^d$  & O6-O7& 7560 \\ 
 17441-2910$^b$ & 10.29 & 3.42 & 2.41 & 30 & O5.5 & 480 \\ 
 18079-1756$^c$ & 11.09 & 2.67 & 1.67 & 74 & B    & 960 \\  \hline
 \end{tabular}
\end{flushleft}
a. Walsh et al. (1999) \\
b. Porter et al. (1998), \cite{wal97} \\ 
c. Osterloh et al. (1997) \\ 
d. Using the given spectral type and using a distance of 2.2\,kpc (see text) we derive m$_v$\,=\,6, yet
there is no optical counterpart on DSS plates.
\end{table}


\section{Observations and data reduction}

\subsection{Target selection}

\begin{figure} 
 \centering \includegraphics[width=5cm, height=7cm,angle=90]{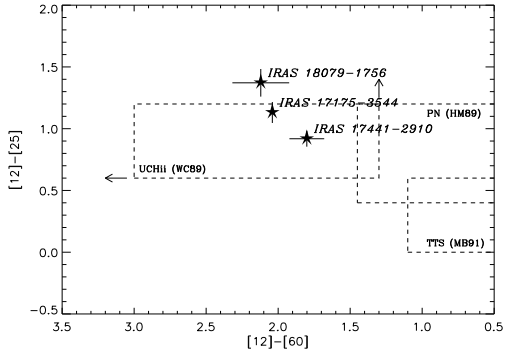}
 \caption{IRAS colour-colour plot of the target objects as filled asterisk with
 errorbars, except the [12]-[60] error of IRAS 17175-3544, for which the 60\,$\mu$m
 IRAS measurement is of bad quality. Also indicated are the colours
 of 1. UC H\,{\sc ii} regions compiled by Wood \& Churchwell (1989), 2.
 Planetary Nebulae (Hughes \& MacLeod, 1989), 3. T\,Tauri stars (Morgan
 \& Bally, 1991). Figure adapted from Osterloh et al. (1997).}
\label{IRAS}
\end{figure}

\begin{figure} 
 \centering \includegraphics[width=5cm, height=7cm,angle=90]{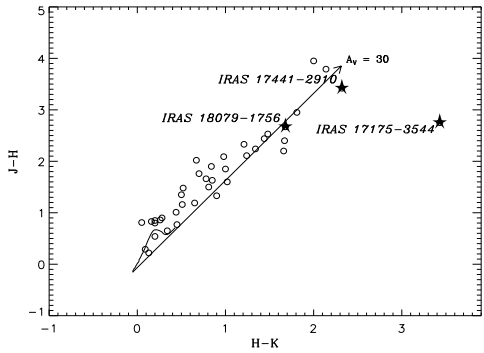}
 \caption{2MASS near infrared colour-colour diagram. Target object are indicated
 again by asterisks which are larger than the errorbars ($\pm$\,0.06\,mag or
 smaller for 2MASS colours, $\pm$\,0.1 for \cite{tap96}). The short curve shows
 colours for main sequence stars (Koornneef 1983). The long straight arrow
 corresponds to displacements of an early type object suffering an extinction
 equal to $\rm A_{V}=30$. Displacement to the right of this line indicates
 a near-infrared excess. The small open circles are 2MASS detected objects,
 within 1\arcmin~ radius from the targets. Note the strong extinction towards
 our targets.}
\label{nircc}
\end{figure}

Targets for this study were selected from recent literature on UC H\,{\sc ii}
regions and IRAS-selected catalogues of massive YSOs (see Table 1).  The main
criteria were that such targets would be sufficiently bright in the H-band to be
accessible for medium resolution ISAAC spectroscopy in short exposure times, and
that they should be present in the publicly available 2MASS dataset, to avoid the
need for prior imaging. A further important criterion was that the 2MASS point
sources should have no optical counterparts in the GSC2.2 and USNO
catalogues. The three selected targets are listed in Table 1 together with 2MASS
magnitudes and colours, A$_{\rm v}$, and indirect spectral type estimates from
the given references. The
mid-infrared fluxes are shown in the IRAS colour-colour
diagram (Fig.\,\ref{IRAS}), where different locations occupied by sources of various types
are indicated. IRAS\,17175-3544 is too faint in the J-band 
for 2MASS to give an accurate magnitude in this band. However, this source shows a strong near infrared excess 
as shown in Fig.\,\ref{nircc} where its JHK colours were derived instead from 
the deeper imaging of Tapia et al. (1996).

\begin{figure}
\centering \includegraphics[bb=96 186 517 605,width=5cm,clip]{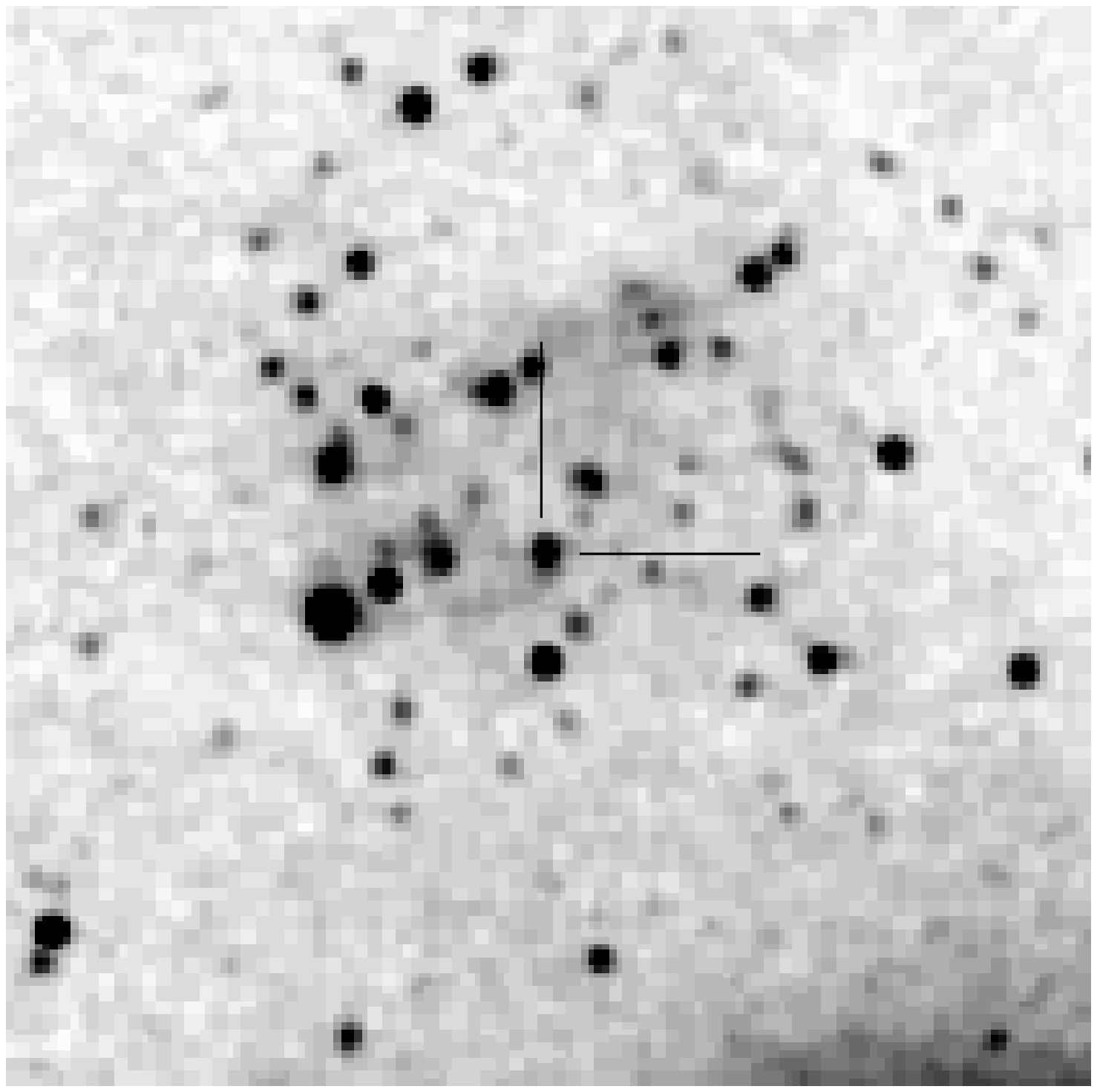}
\centering \includegraphics[bb=95 206 517 588,width=5cm,clip]{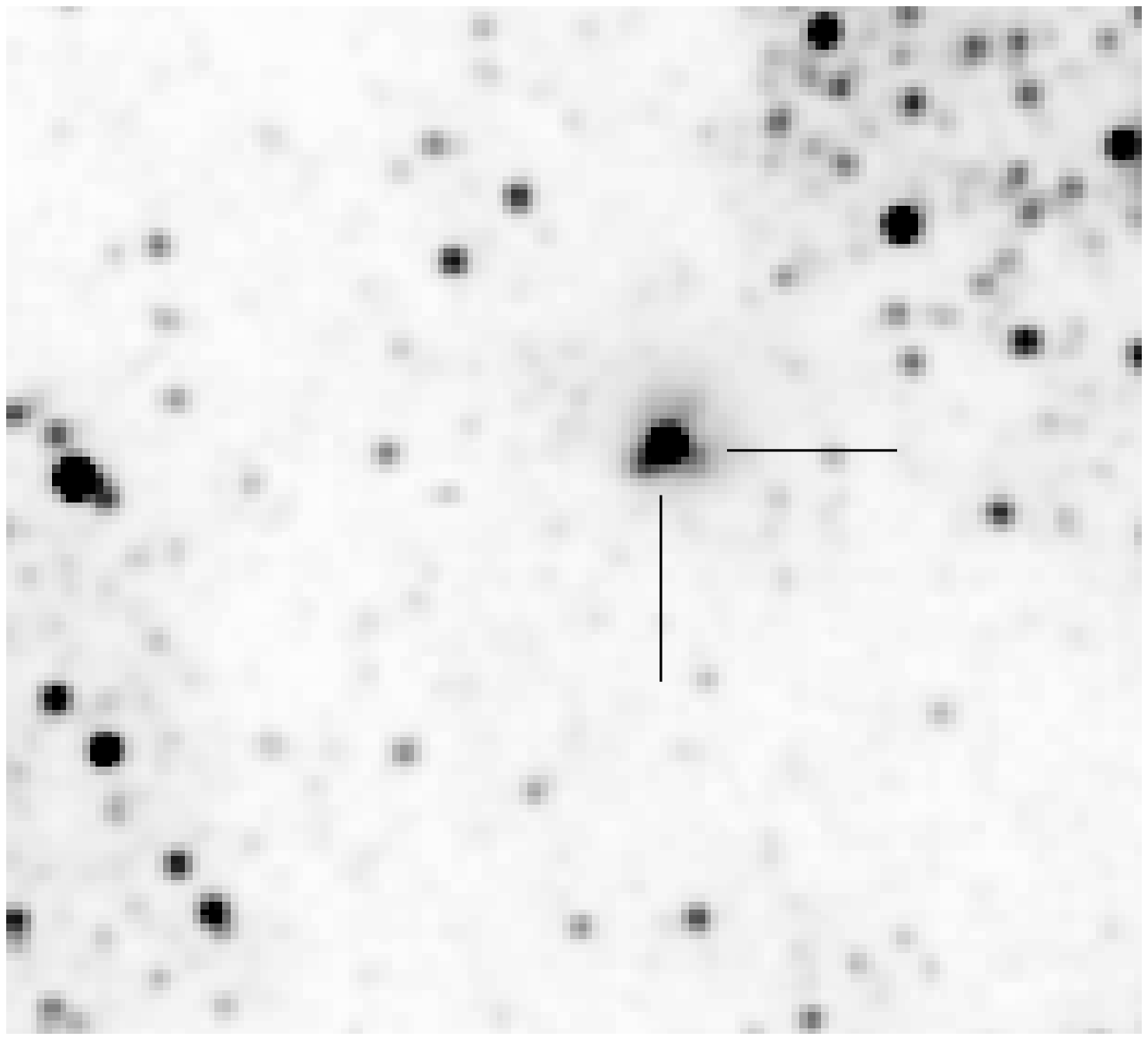}
\centering \includegraphics[bb=94 188 515 606,width=5cm,clip]{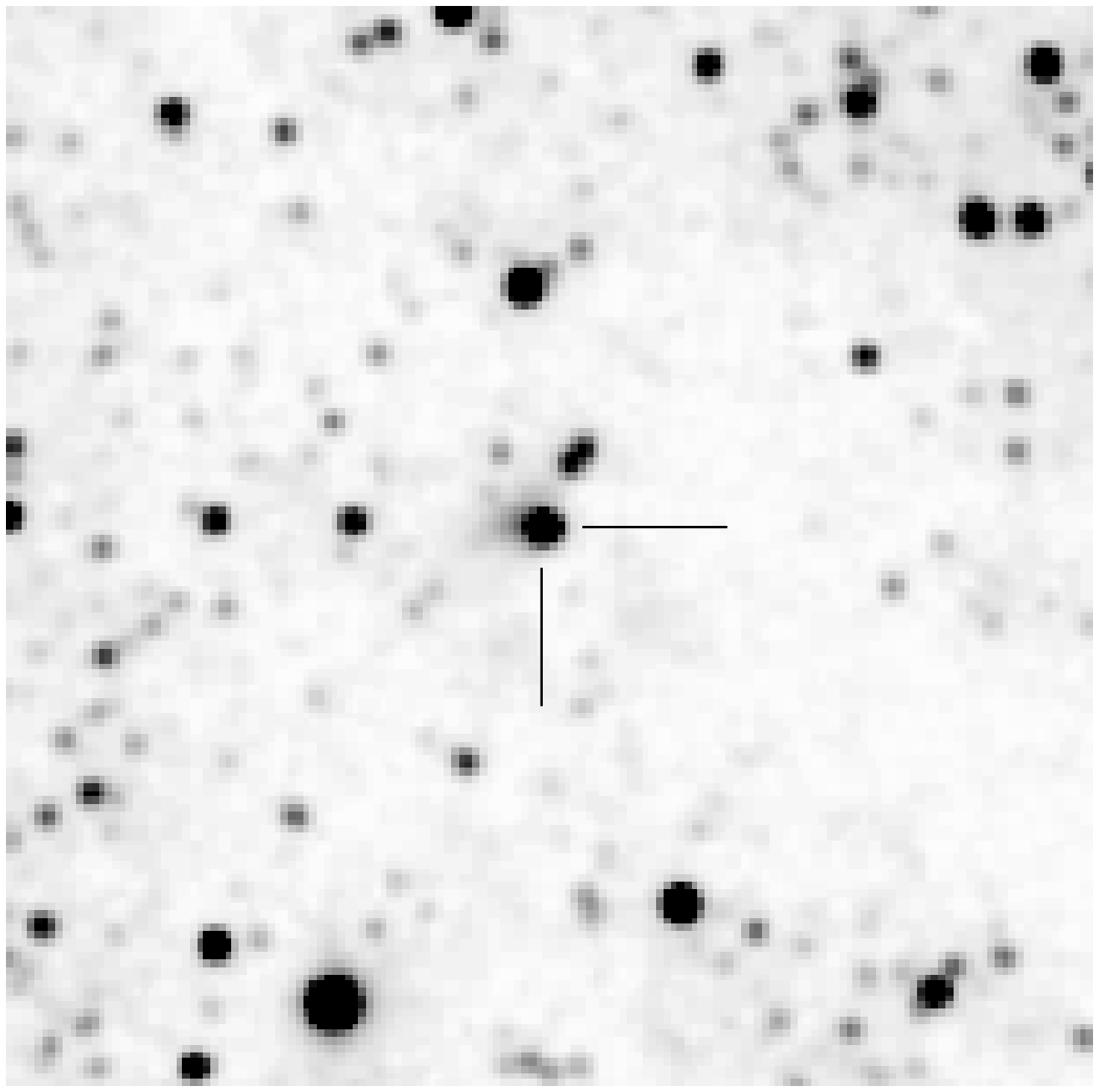}
\caption{2MASS H-band image of IRAS 17175-3544 (top), IRAS 17441-2910 (centre)
and  IRAS 18079-1756 (bottom). North is up, East to the
left. The field size is 2.5\arcmin~ square.}
\label{h17175}
\end{figure}

Fig.~\ref{h17175} shows the 2MASS H-band
images, covering 152\arcsec~ square (the FOV of ISAAC) for each target.
Fig. ~\ref{h17175} can be usefully compared to Fig. 19 of Walsh et al. (1999)
for IRAS\,17175-3544, who show a K-band image overlaid with the positions of
methanol and OH maser sites and 8.64 GHz radio continuum emission, which are
shown to be coincident with the near-infrared source.  Furthermore, L-band
imaging of IRAS\,17175-3544 by the same authors indicates that the source is
extremely bright and red (L=3.8, {\it K-L}=4.7).  In the cases of IRAS\,17441-2910 
and IRAS\,18079-1756, 2MASS H-band data indicate that the bright
sources shown in Fig.~\ref{h17175} lie within 2--3\arcsec~
of the IRAS positions, allowing unambiguous identification of the targets.

\begin{figure}
\centering \includegraphics[height=5cm,width=5cm]{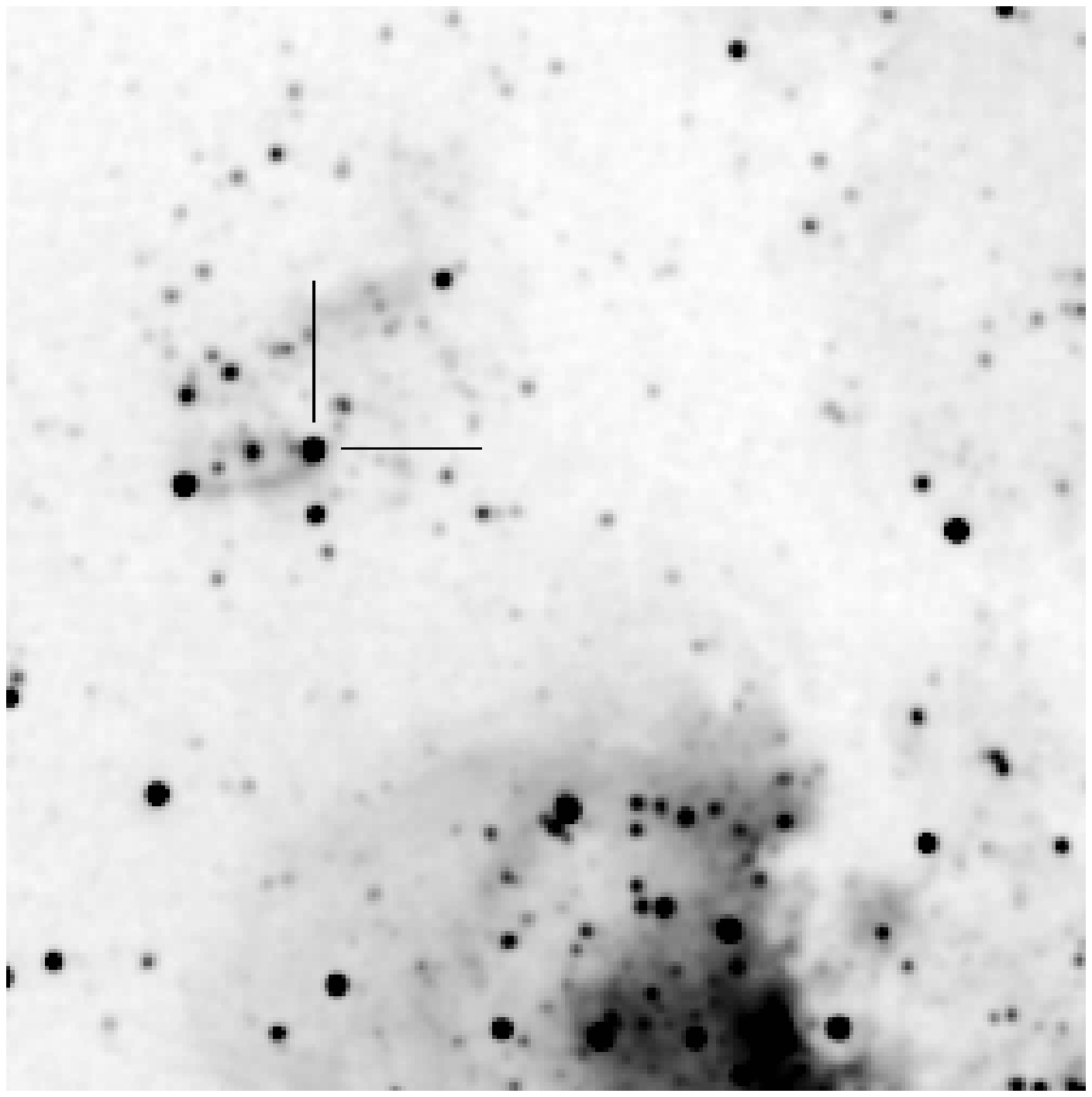}
\centering \includegraphics[height=5cm,width=5cm]{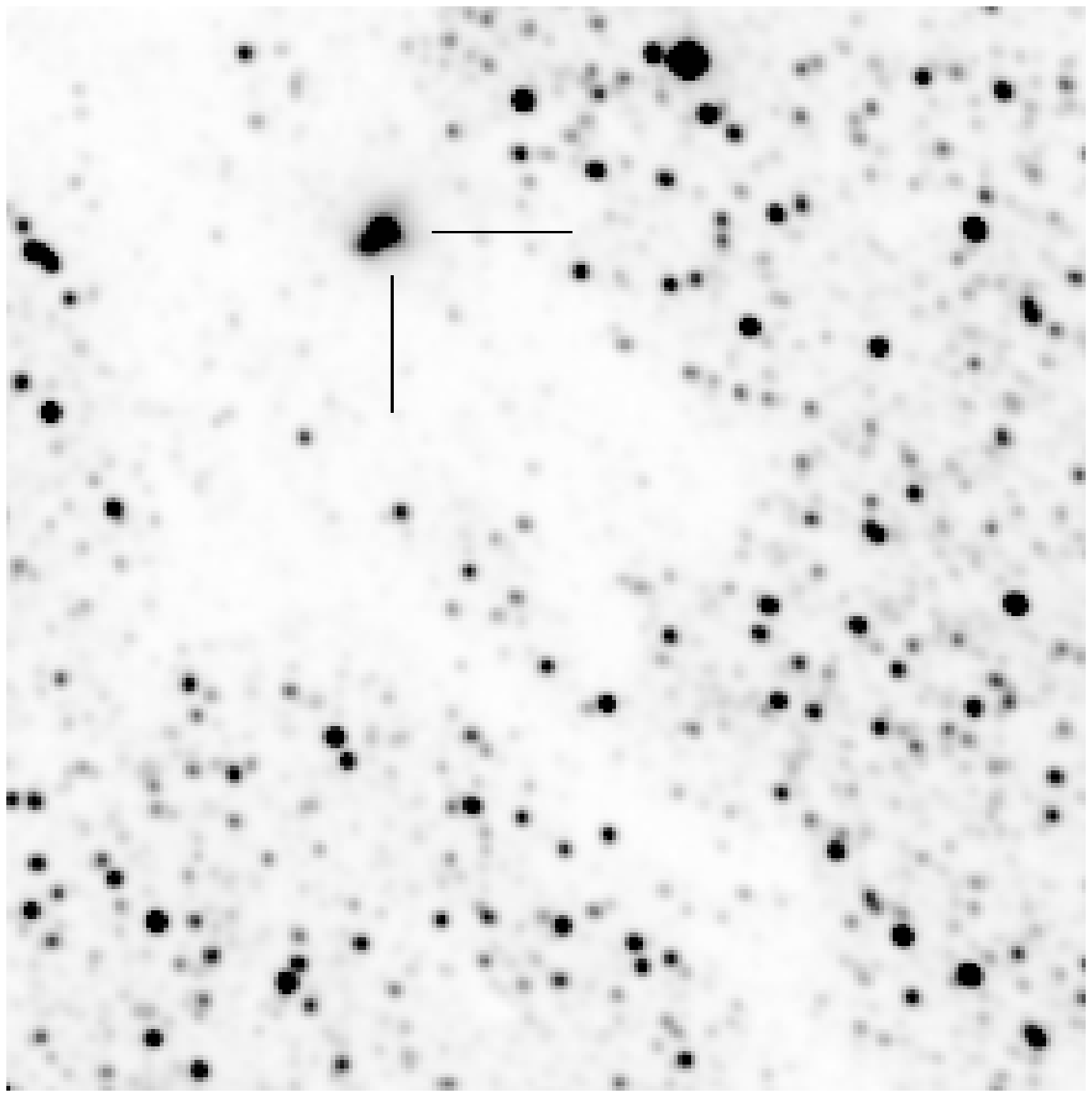}
\centering \includegraphics[height=5cm,width=5cm]{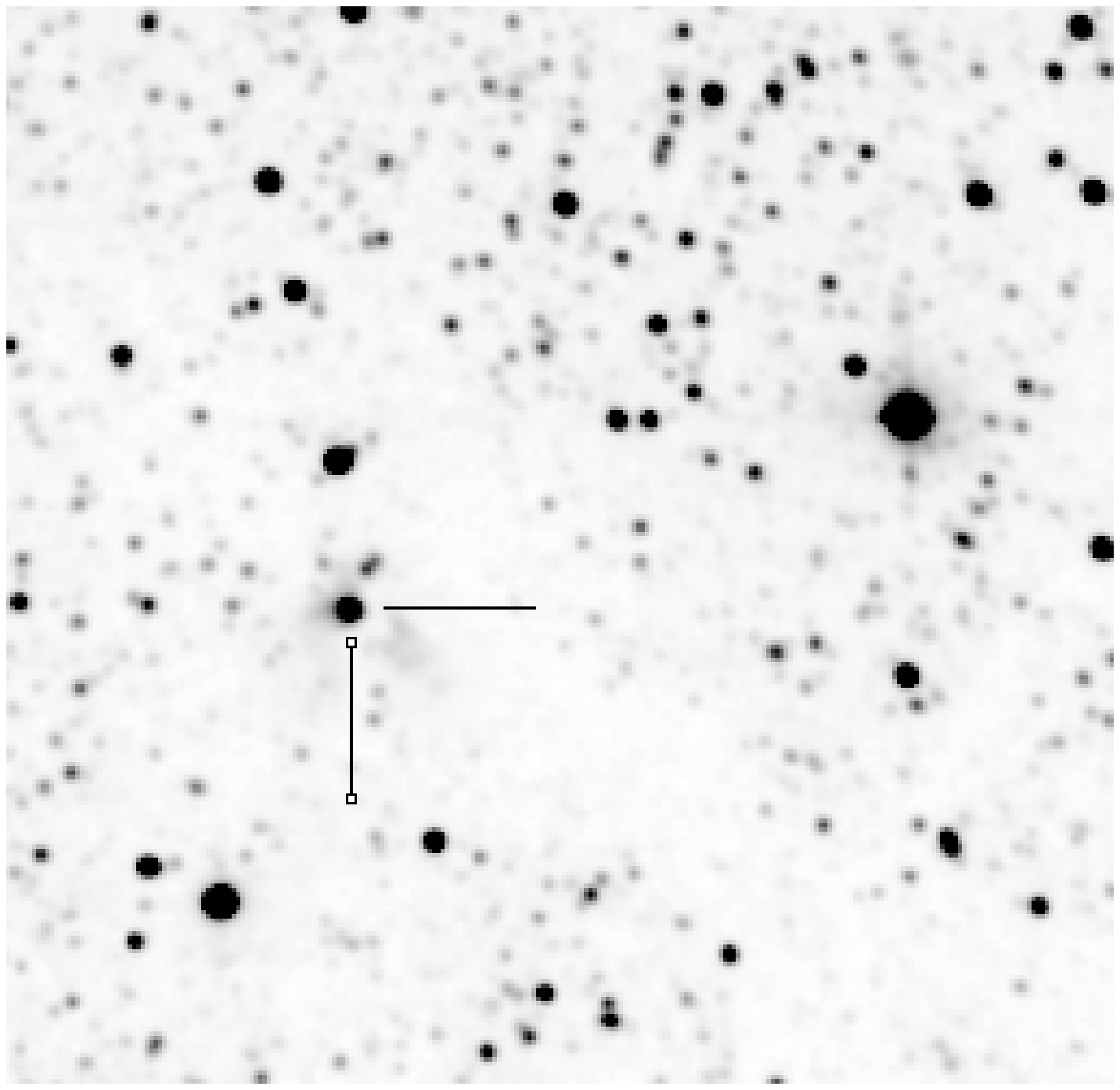}
\caption{2MASS K-band image of IRAS 17175-3544 (top), IRAS 17441-2910 (centre)
and IRAS 18079-1756 (bottom) . North is up, East to the
left. The field size is 4\arcmin~ square. In the top panel (IRAS\,17175-3544), the North-Eastern part of the image is
NGC6334\,I; in the South-Western part is the star forming region NGC6334\,II
(see text). Note the strong extinction, obvious
from the small number of stars near the target objects, which are indicated.}
\label{k17175}
\end{figure}

The presence of nebulosity associated with all targets is clearly shown in the
H-band images and even more obvious in the K-band (see
Fig.~\ref{k17175}, which uses the same intensity
scales as the equivalent H-band images).  This strengthens the identification of
the sources in the 2MASS data with the IRAS sources given in the literature.
The very bright source IRAS\,17441-2910 appears to be multiple, but manipulation
of the image indicates that there is one single, major source for which the
point spread function in the 2MASS H-band image is symmetric. Therefore it was
included as a target under the assumption that this one source would contribute
by far the greater flux to the H-band spectrum.

\subsection{Data extraction}

Observations were carried out at VLT/UT1 (Antu) with the ISAAC spectrograph, in
SW mode at medium resolution (R$\sim$5000), in service mode, on 2002 May 9 and
10. The total integration time for each object is given in Table 1 and totals
150 min.  A central wavelength of 1.71\,$\mu$m was selected to provide a
wavelength range of $\sim$1.67 -- 1.75\,$\mu$m, covering lines diagnostic of an OB
spectral type, most importantly H\,{\sc i} Br 10 1.7362\,$\mu$m, Br 11
1.6811\,$\mu$m, He\,{\sc i} 1.7007\,$\mu$m and He\,{\sc ii} 1.693\,$\mu$m. Data
reduction was performed with the ISAAC pipeline package ECLIPSE (version
4.1.2). Firstly, all data and calibration frames were "de-ghosted" using the
{\it ghost} routine. Next, master flat fields were created for each night using
the {\it sw\_spflat} routine. Co-added spectrum images were then created using
the {\it sw\_spjitter} routine. Full details of the ISAAC reduction pipeline can
be found in \cite{ami02}.

Spectra were then extracted using the standard {\sc apall} routine in {\sc
iraf}\footnote{{\sc iraf} is distributed by the National Optical Astronomy
Observatories, which is operated by the Association of Universities for Research
in Astronomy, Inc. (AURA) under cooperative agreement with the National Science
Foundation.}, and wavelength calibrated using the sky OH emission lines found
throughout the spectra. In all cases $\sim$17 OH lines were used, yielding RMS
values on the fits to the dispersion of $<$0.1\,\AA.

Telluric divisor stars were selected at the time the observations were made, and
were typically observed to within 0.05 airmasses of the science objects.
Telluric standards were reduced, extracted and wavelength calibrated in exactly
the same way as the science data. As the divisor stars are all of spectral type
B, Br 11, Br 10 and He\,{\sc i} profiles were estimated by eye and directly 
fitted out of the standard spectra, and pure telluric spectra created in these regions by dividing the
spectrum by the fit. These sections were then spliced into the rest of the
standard star spectra (assumed free of photospheric lines) to create normalised
pure telluric spectra, one derived from each standard star observation.

\begin{figure}
\centering \resizebox{\hsize}{!}{\includegraphics[bb= 0 80 543 790]{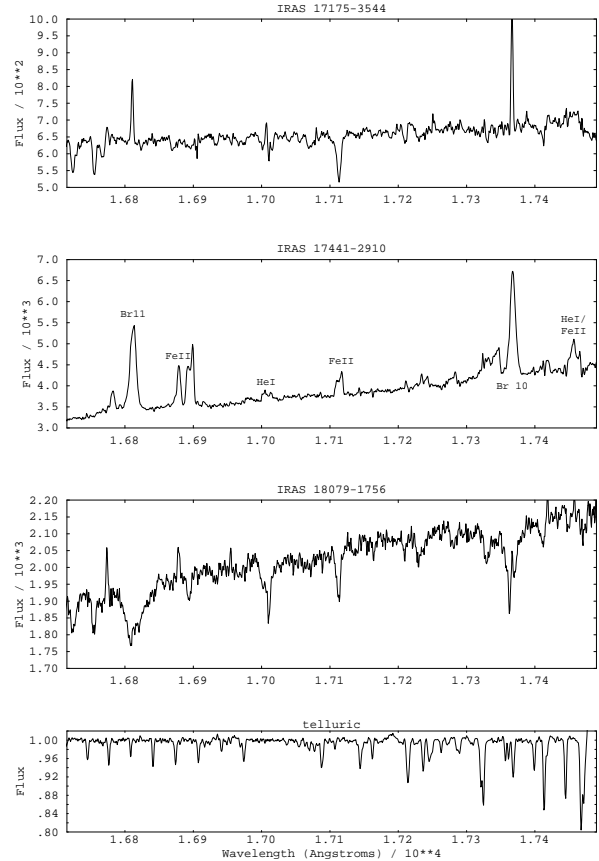}}
\caption{H-band spectra of the target objects, covering the complete wavelength
range 16715 -- 17490\AA, and corrected for telluric absorptions (except near 1.70\,$\mu$m: see text). 
Upper large panel: IRAS\,17175-3544.
Central: IRAS\,17441-2910. Lower: IRAS\,18079-1756. Note that the flux scale (counts) is different
for IRAS\,17441-2910, in order to show the strong emission lines. A normalised template
telluric spectrum used is shown in the bottom smaller panel.}
\label{spec}
\end{figure}

\begin{figure}
\centering \resizebox{\hsize}{!}{\includegraphics[bb= 0 400 543 790]{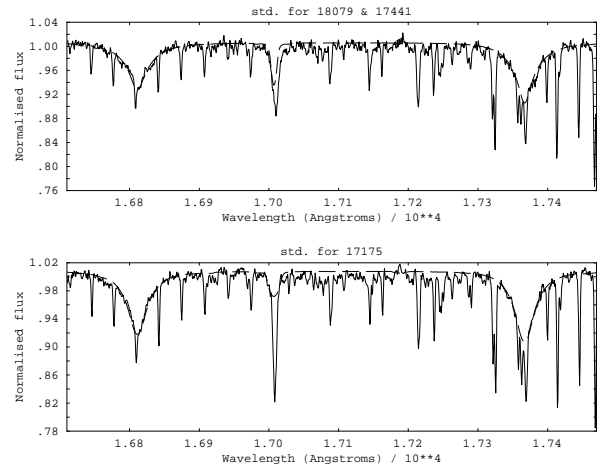}}
\caption{Normalised spectra of telluric standards for IRAS\,17175-3544 (top)
and IRAS\,18079-1756 (bottom) with best fit model spectra (dashed lines).
See text for the model parameters.}
\label{model}
\end{figure}

One problem is that in all standard star spectra He\,{\sc i} is itself blended
with a telluric line, too closely for the components to be separated. Therefore
the whole He\,{\sc i} + telluric profile had to be fit, and replaced by a flat
`continuum' region. Hence the region around He\,{\sc i} remains uncorrected for
telluric absorption in all target spectra shown in Fig.~\ref{spec}. The problem is dealt with further in
Sect. 4, where we have used synthetic spectra to model the telluric standards (see Fig.~\ref{model}).

Object spectra were divided by those of the appropriate normalised standard
(after fitting out photospheric lines) using the {\sc iraf} package {\sc
telluric}. This method retains the same flux scale (number of counts) and
spectral slope as the uncorrected spectra. The signal-to-noise ratio (S/N) of the
final spectra is $\geq$100
and in the case of the very bright target IRAS\,17441-2910, it approaches 150.

\section{Background for individual objects}

The three target objects are located in heavily obscured regions in the Galactic
plane. Strong extinction in the respective areas is especially obvious from
2MASS K-band images (Fig.\,\ref{k17175}). This
results in a sharp drop in 2MASS K-band detections within a few
arcminutes of the target objects.  The obscuring material appears to have a
filamentary structure, especially visible near IRAS\,17441-2910 and
IRAS\,18079-1756. Neither of these two objects are located within a known 
massive star forming region or giant molecular cloud. 

By contrast, IRAS\,17175-3544 lies in the North-Eastern part of the extended molecular
cloud complex NGC6334 (photometric distance 1.76\,kpc, Neckel 1978). This star
forming region is known as NGC\,6334 I (Emerson et al. 1973; McBreen et
al. 1979) and contains three IR sources IRS-I 1, IRS-I 2 and IRS-I 4 (Harvey \&
Gatley 1983). IRS-I 1 and IRS-I 2 are within a few arcsec of each other, and both
have been proposed to be the driving source of a high-velocity ($\rm 70\,km\,s^{-1}$) CO
bipolar outflow (see Persi et al.  1996).

IRAS\,17175-3544 itself appears to be associated with IRS-I 1 (see Fig.\,19 of Walsh et al. 
(1999) and Persi et al. 1996). High resolution near-IR imaging showed IRS-I 1 to be
complex, consisting of at least 4 components (Persi et al. 1996). The nearby (6\arcsec)
source IRS-I 2 is a mid-IR source (detected at 20 and 30\,$\mu$m). Whether or not this
source contributes to the H-band spectrum is determined by its extension and position 
with respect to the slit, but no source brighter than K\,$\sim$\,17.5 is detected at the
position of IRS-I 2 by Persi et al. (1996), so we consider such a contribution unlikely. 
This source could however
contribute to the IRAS fluxes.

Further evidence for the massive YSO nature of at least two of the objects
discussed in this work, namely IRAS\,17175-3544 and IRAS\,18079-1756, is
that both have IRAS low resolution (8 -- 23\,$\mu$m) spectra classified as type H by Kwok et al. (1997). These authors state that the vast majority of IRAS LRS
spectra of this type, exhibiting very red continua, are associated with H\,{\sc ii} regions. 

Another noteworthy feature concerning IRAS\,17175-3544 is that it is
surrounded by an embedded stellar cluster (centred on IRS-I 2), which was
detected in the K-band by Straw et al. (1989). The cluster is clearly visible in
Fig.\,\ref{k17175}.  Deeper imaging by Tapia et al. (1996) yielded a stellar density
estimate of about $\rm 1200\,pc^{-3}$, and cluster size of 0.6 pc (for $\rm D=1.74\,kpc$).

In addition to the cluster near IRAS\,17175-3544, our acquisition images
revealed that IRAS\,18079-1756 is also accompanied by a embedded
cluster, which the 2MASS image in Fig.\,\ref{h17175} is too shallow to fully reveal. 
It is not clear whether there is an increase
in stellar density near the third source IRAS\,17441-2910.

Finally, apart from the location in an obscured region and the presence
of a cluster (near at least two sources), a third common 
characteristic of the target sources is that all are associated with maser emission 
and/or are known to be outflow sources, as has already been mentioned for 
IRAS\,17175-3544. IRAS\,17441-2910 shows evidence for OH masering (Caswell 1998),
while IRAS\,18079-1756 displays $\rm H_{2}O$ maser emission located at $\sim$12\arcsec~
distance and is known to be a CO outflow source (Osterloh et al. 1997). To
conclude, all three sources show all the characteristics of young, actively forming massive stars.

\begin{figure}
\centering \resizebox{\hsize}{!}{\includegraphics[bb= 0 600 543 790]{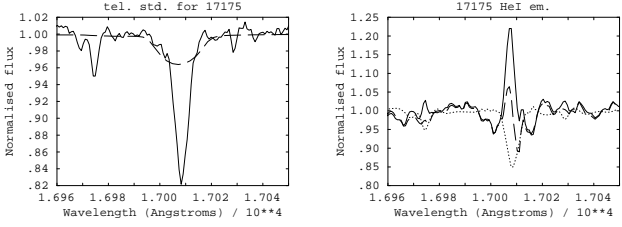}}
\caption{Left panel: Comparison of the region of the He\,{\sc i}\,+\,$\oplus$
line in the telluric standard for IRAS 17175-3544 with the best-fit synthetic spectrum. 
(dashed line, see Fig.~\ref{model} for the complete synthesis). Right panel: The He\,{\sc i} line in IRAS 17175-3544 (dashed line) together with a pure telluric line
derived by division of the two spectra in the left panel (dotted line). The heavy line shows
the results of the division, clearly showing the
narrow He\,{\sc i} emission in the IRAS 17175-3544 spectrum.}
\label{spec5}
\end{figure}

\begin{figure}
\centering \resizebox{\hsize}{!}{\includegraphics[bb= 0 600 543 790]{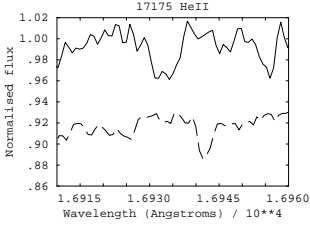}}
\caption{Detection of the He\,{\sc ii} line at $\sim$ 1.693\,$\mu$m in
absorption in IRAS 17175-3544. The telluric spectrum in the same region (offset
for clarity) is also shown (dashed line) and suggests the He\,{\sc ii} detection
is not a result of imperfect $\oplus$ cancellation.  }
\label{spec6}
\end{figure}

\section{Results from individual spectra}

\subsection{IRAS 17175-3544}

Fig.~\ref{spec} shows the observed H-band spectrum of IRAS\,17175-3544 (upper large panel).
As has
already been discussed, Walsh et al. (1999) find maser and radio continuum
sources exactly coincident with the near-IR source discussed here.  IRAS\,17175-3544 
is one of the few sources for which Walsh et al. (1999) can
unambiguously derive a spectral type. As a result of a methanol maser emission
survey to identify UC H\,{\sc ii} regions from an IRAS-selected sample,
\cite{wal97} calculated total luminosities and spectral types directly from the
IRAS fluxes and the kinematic distances given by the maser emission. However,
the large size of the IRAS beam ($\sim$1\arcmin), may encompass a number of
emitting regions, so the IRAS-based spectral types cannot be definitive.

In the case of IRAS\,17175-3544, though, extrapolation of the near-IR SED to the
far-IR shows that the near-IR counterpart will be the one {\it dominant} star in
the field. Using a kinematic distance of 2.2 kpc, Walsh et al. (1999) find O6 --
O7 (from a total luminosity of 12.1$\times$10$^4$\,L\sun).

The H-band spectrum does not slope as steeply to the red as the other two
objects in the sample, perhaps suggesting it is less embedded; yet the infrared
excess is by far the largest of this sample. Narrow, relatively weak emission from
H\,{\sc i} Br 10 and Br 11 is clearly seen. In Fig.~\ref{spec}, the region
around the He\,{\sc i} line at 1.7007\,$\mu$m is not corrected for telluric lines,
for reasons already discussed. However, there is certainly a suggestion of
He\,{\sc i} emission. This is clarified in Fig.~\ref{spec5}, where in order to perform telluric correction 
in this region, we have fitted the complete telluric standard spectrum with a synthetic spectrum.
For the standard for IRAS 17175-3544, we adopt T$_{\rm eff}$\,=\,20000\,K, log $g$\,=\,3.5 and vsin$i$\,=\,200
km\,s$^{-1}$ (see also Fig.~\ref{model}). The He\,{\sc i} line for these parameters is shown in the left panel of Fig.~\ref{spec5} (dashed line)
together with the blended He\,{\sc i} + telluric profile in the standard star. After division, a pure telluric line is
obtained (Fig.~\ref{spec5}, right panel, dotted line). Using this profile, the He\,{\sc i} region of IRAS 17175-3544
was corrected for telluric absorption, yielding a clear emission line rising to $\geq$\,20\% above the continuum level. 
Using this line, together with the narrow H\,{\sc i}
emission, we derive an LSR radial velocity of +15.9\,$\pm$\,1.5\,km\,s$^{-1}$
for this object. This is somewhat different from the methanol maser
emission velocity given by \cite{wal97}~(--10\,km\,s$^{-1}$) and could
indicate that the atomic and molecular emission do not arise in precisely the
same circumstellar region. However we note that the velocity resolution of our data is insufficient to press this point.
It is clear, however, from inspection of the 2D spectrum images for this object, that the H and He emission lines
are extended spatially, and hence are of nebular origin.  

By themselves, emission lines of H and He suggest an early-type object, but do
not allow derivation of a spectral type. We may speculate that, as the emission
is much weaker than in IRAS\,17441-2910, either the exciting central object is
not as luminous in this case, or perhaps the object is in an `intermediate'
stage of evolution, displaying nebular emission but not the powerful
circumstellar emission of IRAS\,17441-2910,
nor resembling a main sequence type spectrum like IRAS\,18079-1756 (see below).

As this object has previously been characterised as O6 -- O7, we have searched
the He\,{\sc ii} region carefully. Fig.~\ref{spec6} shows this region in the
telluric corrected spectrum, with the telluric standard star also plotted as a
dashed line. The feature in the IRAS\,17175-3544 spectrum at $\sim$1.693\,$\mu$m
does not appear to be an artifact of the telluric correction process, since the
standard star spectrum is featureless across its profile.

We identify this feature as He\,{\sc ii} absorption. The equivalent width is
close to 300\,m\AA, compatible with the measurements of \cite{han98}: in their
small sample, they measure 0.5$\pm$0.2\,\AA~ for their latest O-dwarf
(HD48279, O8V) and 0.3$\pm$0.2\,\AA~ for the O7V HD 47839, while they do not detect
He\,{\sc ii} in their O9.5V object.

Moreover, \cite{mey98} include the O7V star HR2546 (15 Mon) in their sample.
Even their best spectrum of this object is rather noisy (S/N$\sim$50), but
clearly exhibits He\,{\sc ii} absorption. Since the spectrum is noisy, and the
continuum level around He\,{\sc ii} uncertain, we have not attempted to plot the
profile in Fig.~\ref{spec6}. However, the equivalent width is in the range 150
-- 400\,m\AA, close to both \cite{han98} and our measurement. Our He\,{\sc ii}
detection therefore supports the previous classification of IRAS 17175-3544 as
O6 -- O7. A caveat, however, is that according to Hanson et al. (1998),
He\,{\sc i} should still be strong in absorption ($\sim$\,1\,\AA) for an O7V type. Such absorption appears to be masked by nebular emission in IRAS\,17175-3544.

That IRAS\,17175-3544 is located at the centre of the complex star-forming region of
NGC6334\,I has already been discussed. In addition to the
H and He features already discussed, it shows absorption features which can be
identified as Mg\,{\sc i} (17113\,\AA) and the Al\,{\sc i} triplet (16723, 16755 and
16768\,\AA). Usually these features are associated with cooler atmospheres,
increasing in strength to K-types (Meyer et al. 1998). Inspection of the
telluric standard spectrum excludes a telluric nature for this lines (unlike the
emission spikes near Br\,11 in IRAS\,18079-1756), nor a badly determined and
subtracted sky spectrum (contamination by other, late-type stars in the slit),
since the slit position angles were carefully chosen to avoid any other stars
along its length.
However, a close examination of the VLT acquisition image for IRAS\,17175-3544 reveals the presence
of a second object, which would have been in the slit at the time the spectrum
was obtained. The two sources are not resolved in the 2MASS image.

These two sources are coincident with IRS1-I\,1E and IRS-I\,1W of Persi et
al. (1996, see also Tapia et al. 1996) where the former source is shown by these
authors to be the reddest (H\,=\,15, H-K\,=\,3.9), dominating the emission
longward of 2\,$\mu$m and coincident with the peak of the 30\,$\mu$m
emission. This source is the OB star responsible for the ionization of the
compact H\,{\sc ii} region and for pumping the nearby H$_2$O and OH masers. The
second source has H\,=\,14.4 and J-H\,=\,2.3, and clearly also contributes to
the H-band spectrum, giving rise to the composite spectrum we observe. The
presence of the Mg\,{\sc i} and Al\,{\sc i} lines, presumed to arise from
IRS1-I\,1W, shows that this source is cooler. We have performed experiments with
synthetic spectra which show that the flux emitted by a T$_{\rm
eff}$\,=\,20000\,K photosphere is only $\sim$\,5 times greater than one of
T$_{\rm eff}$\,=\,5000\,K in the H-band, so only a very few cooler embedded
stars might suffice to cause the appearance of Mg\,{\sc i} and Al\,{\sc i}. In
this scenario, IRS-I\,1W might be a system of less massive embedded objects
associated with IRS-I\,1E, the embedded OB star which gives rise to the H and He
features in our spectrum.

Lastly, it might be argued that the observed spectrum could be that of a
foreground late-type object, with emission lines from the nebular background. We
cannot definitively rule out this possibility using the data presented here, but
would point out that there is no optical counterpart to such a putative
foreground object, even on the very deep SuperCosmos (UKST I-band, \cite{ham01}
and references therein) images.

\subsection{IRAS 17441-2910}

The very high signal-to-noise ($\sim$150) spectrum of this bright (H=10.29) source
is shown in Fig.~\ref{spec} (central large panel). The spectrum exhibits broad, strong
emission lines of H\,{\sc i} Br 10, Br 11 and numerous other species, notably
He\,{\sc i} and [Fe\,{\sc ii}]. The emission lines we have been able to identify are
labelled in Fig.~\ref{spec}. The object is known to exhibit very strong CO emission in the K-band
(\cite{por98}).  The equivalent widths of the Br 11 and Br 10 emissions are
$\sim$6\,\AA~ and $\sim$5.5\,\AA~ respectively, with FWHM $\sim$180\,kms$^{-1}$
 and 160\,kms$^{-1}$.  Porter et al. (1998) find Br$\gamma$ in emission with equivalent
width 12\,\AA.  \cite{wal97} have derived a spectral type of O5.5 from the IRAS
fluxes, together with the kinematic distance given by the methanol maser (6.67~
GHz) emission profile (9.8\,kpc), and assuming the source is a single
star. However, the lack of photospheric absorption lines precludes a spectral
type determination here.

After correction to the LSR velocity, we derive a radial velocity of +44.9\,$\pm$\,4\,km\,s$^{-1}$ using 4 lines labelled in Fig.~\ref{spec} which we are able to deblend accurately: Br 11 16811.11, Br 10 17366.85, Fe\,{\sc ii} 16877.81 and Fe\,{\sc ii} 17115.95 \footnote{Rest wavelengths from www.pa.uky.edu/$\sim$peter/atomic/}. This velocity is somewhat different
than that cited by \cite{wal97}, and again may indicate that the atomic emission
arises in a different region from that producing the methanol maser emission (centred near +23\,km\,s$^{-1}$). However, owing to the large FWHM of the emission lines measured, and consequent difficulties in accurately measuring the line centre, we do not consider this point further.

Kaper et al. (2002) have drawn attention to a significant population of objects
whose K-band spectra also show broad, strong emission lines, up to 100
kms$^{-1}$, two of which also show CO emission. IRAS\,17441-2910 clearly belongs
in this category also. Following the suggestion of \cite{kap02}, IRAS\,17441-2910
appears to be a third object for which strong emission of CO and other
species, together with a significant infared excess (from 2MASS colours), indicate
that the exciting object is surrounded by dense circumstellar material. It is
clear that the strong emission line objects represent massive stars in the
earliest stages of evolution.

\subsection{IRAS 18079-1756}

Previous literature on this object gave L$_{bol}$ $\sim$ 32000\,L\sun~ and A$_{\rm
v}$ = 74 (Osterloh et al. 1997). This bolometric luminosity would correspond
roughly to an early B spectral type. The entire H-band spectrum is shown in
Fig.~\ref{spec} (lower large panel). The plot shows that the continuum slope for
IRAS\,18079-1756 rises to the red across the length of the H-band spectrum by
$\sim$20\%, comparable to IRAS\,17441-2910, suggesting it is indeed deeply
embedded. However, the absence of significant emission, and the general
appearance of the spectrum (similar to an unobscured object of the same type)
suggests that it could be at a somewhat later evolutionary phase.

\begin{figure}
\centering \resizebox{\hsize}{!}{\includegraphics[bb= 0 600 543 790]{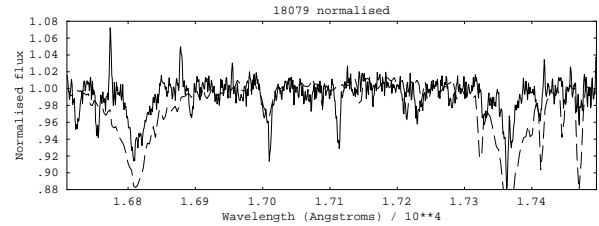}}
\caption{Normalised spectrum of IRAS\,18079-1756 overlain with the H-band
spectrum (dashed line) of the fast rotator HR 5191 (spectral type B3V - see
text). Good agreement in the region of the He\,{\sc i} absorption can be
seen. Mismatches in the strength of the H lines results from filling in of the
line profile by emission. }
\label{spec3}
\end{figure}

\begin{figure}
\centering \resizebox{\hsize}{!}{\includegraphics[bb= 0 600 543 790]{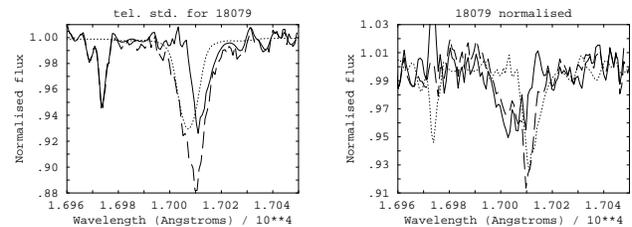}}
\caption{Left panel: The blended He\,{\sc i}\,+\,$\oplus$ profile in the telluric standard spectrum for 
IRAS 18079-1756 (dashed line) together with the adopted model He\,{\sc i} line (dotted line; see text). The 
solid line shows the pure telluric line derived by division. Right panel: This pure telluric line (dotted)
overlain with the uncorrected data for IRAS 18079-1756 (dashed line). The heavy line shows the corrected
He\,{\sc i} profile for this object.}
\label{spec4}
\end{figure}

The spectrum clearly shows broad He\,{\sc i} absorption, which is centred at
$\sim$ 17007\AA, as well as broad hydrogen absorption.
In Fig.~\ref{spec3}, the normalised spectrum of IRAS\,18079-1756 is compared to
an H-band spectrum of HR\,5191 ($\eta$ UMa). This star is one observed by
\cite{mey98}, who give a spectral type of B3V and vsin{\it i} = 205
kms$^{-1}$. The He\,{\sc i} profile is well matched, but the H lines are weaker
than in the standard star, presumably because the H profiles are filled in by
emission. This is supported by inspection of the Br 10 profile, which clearly
shows central emission. The He\,{\sc i} profile is contaminated by a telluric
line, since it was not possible to correct for telluric absorption over a 50\AA~
region across the He\,{\sc i} profile (see Sect. 2.2). Fig.~\ref{spec4} (left
panel, dashed line) shows the blended He + telluric profile in the B-type standard observed
for IRAS\,18079-1756. In the same way as for IRAS\,17175-3544, we have synthesised this
spectrum (see Fig.~\ref{model}) adopting parameters of T$_{\rm eff}$\,=\,25000K, log\,$g$\,=\,4.0 and vsin$i$\,=\,100\,km\,s$^{-1}$ (dotted line). Division yields a pure telluric profile (solid line). In Fig.~\ref{spec4} (right panel), 
this resultant profile (dotted line)
is divided through the uncorrected spectrum for IRAS\,18079-1756 (dashed line) to yield the corrected
He\,{\sc i} profile (heavy line). This profile matches the broad feature in Fig.~\ref{spec3} well, supporting    
the case that the stellar He\,{\sc i} line is the broader feature. Comparing
this profile to the galactic standard yields a very close match between the two
profiles. It is clear that this match strongly supports an early-B spectral type
for IRAS\,18079-1756, and furthermore suggests it is a rapid rotator, as expected
for very young massive YSOs.  However, a rigorous measurement of vsin{\it i}
would require spectra of rather higher resolution.

After telluric correction, the equivalent width of He\,{\sc i} (heavy profile in Fig.~\ref{spec4}) was measured as 600\,$\pm$\,100\,m\AA. Comparison with equivalent
widths tabulated by \cite{han98} against optical spectral type, suggest IRAS\,18079-1756 
is very close to B2, assuming a dwarf luminosity. Comparison with HR\,5191 
suggests B3V. However, strong He\,{\sc i} persists in supergiants to
$\sim$B7. If IRAS 18079-1756 is of lower gravity, which would not be unexpected
by analogy with low-mass embedded protostars (see Greene \& Lada 1996), then a somewhat
later spectral type is possible. The equivalent width of Br 11 is $\sim
$2.0$\pm$0.1\,\AA, which also suggests early B, although because of the possibility of
 emission this is not such a reliable indicator.

Furthermore, He\,{\sc i} 1.700\,$\mu$m remains strong into the O-star regime, as
far as O7, and the Br 11 widths given by \cite{han98} for their late O-stars
show large scatter, varying non-monotonically from 1.6 to 3.0\AA~ between O9.5V
and O7V.  If IRAS\,18079-1756 is as early as O8V, however, it should show
He\,{\sc ii} absorption at 1.693\,$\mu$m with a strength comparable to that of the
observed He\,{\sc i} profile. Such a feature is not seen. \cite{han98} do not
detect He\,{\sc ii} in their O9.5V object, so in principle IRAS\,18079-1756 could
be as early as this. Using slightly lower resolution (R=2000), comparable S/N
spectra, they give equivalent widths as small as 0.3\,\AA\, reliably. Our upper
limit on the strength of He\,{\sc ii} in IRAS\,18079-1756 is smaller than this
($\sim$150\,m\AA: we marginally detect a feature of this strength close to the
expected He\,{\sc ii} wavelength, but it is compatible with the
noise). Therefore, it is not possible to say definitively that IRAS\,18079-1756
is not an embedded very late O-type photosphere. Nevertheless, since we do not
detect He\,{\sc ii} with certainty, we adopt a spectral type of B2-B3, provided
the luminosity class is nearer that of a dwarf than a supergiant, which seems
likely.

As in the case of IRAS\,17175-3544, IRAS\,18079-1756 also shows a composite
spectrum, with the Al\,{\sc i} and Mg\,{\sc i} lines. The clear presence of a
cluster around the latter object suggests that this spectrum may be composite
for similar reasons to IRAS\,17175-3544; namely the existence of a small cluster
of unresolved low-mass embedded objects around or associated closely with the
massive star. For IRAS\,18079-1756, there is no suggestion of a second object in
the slit which can be identified with a known object, unlike the case of
IRAS\,17175-3544 (IRS1-I\,1W). The acquisition image PSF exhibits some
asymmetry, but how many components it contains is unclear. Once again, we note
that if the Mg\,{\sc i} and Al\,{\sc i} lines are to be explained by a chance
superposition of a late-type foreground object, there is no optical counterpart
at the correct position on the SuperCosmos UKST I-band image.

\section{Discussion}

\subsection{The nature of the three sources}

Our sample of three embedded massive YSOs in UC H\,{\sc ii} regions has
uncovered a variety of objects which may represent different stages of the
earliest evolution of massive stars. 

IRAS\,17441-2910 shows broad, strong emission lines, and is known to show strong
CO emission, indicating the presence of dense circumstellar material. While the
intrinsic near-infrared excess emission is marginal compared to that of
IRAS\,17175-3544, the presence of emission lines suggests mass accretion and/or
outflow processes. This object appears to be at the earliest evolutionary
phase of the three candidate massive YSOs.

IRAS 17175-3544 is in some ways the most enigmatic source, since it has a flat
continuum, but the largest near-IR excess of all. Its association with IRS-I 1
in the molecular cloud complex NGC\,6334 (see Sect. 3) supports a young
evolutionary status. The nebular emission detected (Br 10, Br 11, He\,{\sc i})
is narrower and weaker than in IRAS\,17441-2910, but still clear. If the rather
flat continuum of IRAS\,17175-3544 is taken to mean it is relatively less
embedded than IRAS\,17441-2910, the infrared excess is not explained. Instead,
the weakness of the emission lines compared to IRAS\,17441-2910 may be indirect
evidence that IRAS\,17175-3544 is less luminous than IRAS\,17441-2910, and that
the spectral type of the latter is earlier than O6 -- O7, which is supported by
the previous derivation of a spectral type of O5.5 (see Table 1). Note however,
that the broad emission in IRAS\,17441-2910 is most likely of circumstellar
rather than nebular origin. We propose that IRAS\,17175-3544 may be at an
equally early stage of evolution as IRAS\,17441-2910.

Finally, the spectrum of IRAS\,18079-1756 resembles a main sequence
B-star, lacking direct spectral evidence for circumstellar material. The absence
of near-infrared emission line features suggests a later evolutionary stage than
the other two objects. However, the continuum does slope to the red, rising by
$\sim$\,20\% over the 750\,\AA~ range covered by these spectra, similar to
the slope exhibited by IRAS\,17441-2910. Perhaps tellingly, IRAS\,18079-1756
shows little near-infrared excess, while IRAS\,17441-2910 does.
This supports the hypothesis that IRAS\,18079-1756 is more evolved but still
obscured, possibly predominantly by intracluster dust.

While this would explain the slope of the continuum and the location of the source in Fig.~\ref{nircc},  
the association of the source with a UC H\,{\sc ii} region and H$_2$O masering, and its identification as a CO outflow
source by Osterloh et al. (1997), lend weight to the suggestion that it is an embedded (that is, still 
surrounded by circumstellar material) YSO.

We note also that the very presence of emission lines does not confirm
these sources are massive embedded YSOs; OB supergiants such as P Cygni also exhibit emission lines in the near-infrared (Hanson et al. 1998), but such objects
are not generally optically obscured, nor associated with masering phenomena.

\subsection{Spectral types from H-band data}

Where H and He absorption lines are detected, it is possible to obtain unambiguous
spectral types, by comparison to galactic standards. He\,{\sc ii} absorption is
detected in IRAS\,17175-3544, in close quantitative agreement with the spectral
type derived from IRAS fluxes, O6 -- O7. From the presence of strong He\,{\sc i}
and the absence of detectable He\,{\sc ii} in IRAS\,18079-1756, we also confirm a
spectral type of B3, with considerable accuracy ($\pm$ $\sim$3 subclasses).
These close agreements strongly suggest that near-infrared spectral types could
be derived for many of the known UC H\,{\sc ii} regions for which far-infrared
spectral types, derived from bolometric luminosities, are ambiguous (e.g. Walsh
et al. 1997).

In a very recent paper, \cite{han02} have performed a large survey of UC H\,{\sc
ii} regions in the K-band, including spectroscopy at R=1200. Their main finding
relevant here is that $\sim$ 50\% of their radio-selected sample were detectable
by Br$\gamma$ emission, but of those, only for 5 -- 10\% was it possible to
detect photospheric features of the ionizing star, and thus derive spectral type
and hence effective temperature directly. However, in the K-band spectroscopic
survey of \cite{kap02}, at R$\sim$8000, these authors detect characteristic
K-band OB star features (emission and absorption) in the majority of their
sample of 75, of which $\sim$ $\frac{1}{3}$ are found to be late-type foreground
contaminants, and around 20 are found to be extremely strong Br$\gamma$ emitters
with obvious resemblances to IRAS\,17441-2910, as discussed above. This suggests
that if sufficiently high resolution and high signal-to-noise spectra
(R$\geq$5000, S/N$\geq$100) are used in either the H or K band, the photospheres
of massive YSOs, at very early stages of evolution, could be readily
detected, and absorption lines analysed. Such spectra would be amenable to
similar techniques as those common in optical studies, in particular the
derivation of gravities log {\it g} and projected rotational velocities vsin{\it
i}, both of which may be expected to differ systematically from that observed on
the main sequence. Finally, given suitable line(s) present in the H-band, the detection of a surface enhancement of
$^{14}$N in young massive stars suggests an observational test of the
coalescence formation scenario, by analogy with O-type blue stragglers (\cite{ken95}). 

\section{Conclusions}

   \begin{enumerate}

{\item We have used high signal-to-noise, intermediate resolution H-band
spectroscopy to probe the central, ionizing stars of three UC H\,{\sc ii}
regions. In two of the objects, IRAS\,18079-1756 and IRAS\,17175-3544, we detect
the photosphere directly {\it via} absorption lines, allowing the derivation of
spectral types which are in close agreement with those found by far-infrared
observations.

\item In a third object, we detect no absorption, but a rich emission line
spectrum. We identify this object, IRAS\,17441-2910, as an additional member of the
small subset of massive YSOs displaying strong emission (including CO) and which
are likely to be at an extremely early evolutionary stage, surrounded by dense
circumstellar material. The large infrared excess of IRAS\,17175-3544, coupled
with the observation of emission lines, also places it in this category.

\item IRAS\,17175-3544 and IRAS\,18079-1756 exhibit composite spectra, including
absorption lines of neutral metals normally observed in cool stars, as well as H
and He lines arising in the OB photospheres of or nebulosity associated with the
targets themselves. Both these objects are associated with small clusters, as
observed in our acquisition images. For IRAS\,17175-3544, we identify the source
of the metal lines as IRS-I\,1W (Persi et al. 1996) and suggest that this object
is a system of low mass embedded objects associated with the massive source
itself. For IRAS\,18079-1756 we tentatively suggest a similar scenario, although
there is no direct evidence from imaging. In neither case can the spectra
presented here rule out a chance superposition of a foreground late-type object
causing the composite spectra, but if such a foreground star exists, it does not
appear on SuperCosmos UKST I-band images of the two near-infrared sources.

\item H or K-band spectra at a comparable or slightly higher resolution to those
in this study, with high signal-to-noise ratio, may be amenable to quantitative
analysis; e.g. derivation of projected rotational velocities and perhaps surface
gravities, in the same way as for optical spectra. This will open up a new
method to study massive YSOs at the earliest stages of their formation. Both
bandpasses are likely to be useful: sources will be brightest in the K-band,
where CO is also observed, but the H-band promises greatly reduced veiling
and is likely to be a better wavelength region to seek weak metal absorption
lines.  } \end{enumerate}

\begin{acknowledgements}
TRK and WJdeW acknowledge financial assistance from the European Union Research
Training Network `The Formation and Evolution of Young Stellar Clusters'
(RTN1-1999-00436). The authors thank the referee, Lex Kaper, for exhaustive
comments and discussions. This publication makes use of data products from the
Two Micron All Sky Survey, which is a joint project of the University of
Massachusetts and the Infrared Processing and Analysis Center/California
Institute of Technology, funded by the National Aeronautics and Space
Administration and the National Science Foundation. This research has made use
of the SIMBAD database, operated at CDS, Strasbourg, France.
\end{acknowledgements}


\begin{thebibliography}{}
\bibitem[Amico et al. (2002)]{ami02} Amico, P., Cuby, J.G., Devillard, N., Jung, Y., \& Lidman, C. 2002, ISAAC Data Reduction Guide 1.5, ESO
\bibitem[Bonnell et al. 1998]{bon98} Bonnell, I., Bate, M.R., \& Zinnecker, H. 1998, MNRAS, 298, 93
\bibitem[Bronfman et al. 1996]{bro96} Bronfman, L., Nyman, L-\AA., \& May, J. 1996, A\&AS, 115, 81
\bibitem[Campbell et al. 1989]{cam89} Campbell, B., Persson, S.E., \& Matthews, K. 1989, AJ, 98, 643
\bibitem[Caswell 1998]{cas98} Caswell, J.L. 1998, MNRAS, 297, 215
\bibitem[Chan et al. 1996]{cha96} Chan, S.J., Henning, Th., \& Schreyer, K. 1996, A\&AS, 115, 285
\bibitem[Dallier et al. (1996)]{dal96} Dallier, R., Boisson, C., \& Joly, M. 1996, A\&AS, 116, 239
\bibitem[Emerson et al. 1973]{eme73} Emerson, J.P., Jennings, R.E., Moorwood, A.F.M. 1973, ApJ, 184, 401
\bibitem[Greene \& Lada (1996)]{gre96} Greene, T.P., \& Lada, C.J. 1996, AJ, 112, 2184
\bibitem[Hambly et al. (2001)]{ham01} Hambly, N.C., Davenhall, A.C., Irwin, M.J., \& MacGillivray, H.T., 2001, MNRAS, 326, 1315
\bibitem[Hanson et al. (1998)]{han98} Hanson, M.M., Rieke, G.H., \& Luhman, K.L. 1998, AJ, 116, 1915
\bibitem[Hanson et al. (2002)]{han02} Hanson, M.M., Luhman, K.L., Rieke, G.H. 2002, ApJS, 138, 35
\bibitem[Harvey \& Gatley 1983]{har83} Harvey, P.M., \& Gatley, I. 1983, ApJ, 269, 613
\bibitem[Hughes \& MacLeod 1989]{hug89} Hughes, V.A., \& MacLeod, G.C. 1989, AJ, 97, 786
\bibitem[Ishii et al. 2001]{ish01} Ishii, M., Nagata, T., Sato, S., Yoa, Y., Jiang, Z., \& Nakaya, H. 2001, AJ, 121, 3191
\bibitem[Kaper et al. (2002)]{kap02} Kaper, L., Bik, A., Hanson, M.M., \& Com\'{e}ron, F. 2002, in `The Origins of Stars and Planets: The VLT View', ESO Astrophysics Symposia, Springer-Verlag
\bibitem[Kendall et al. 1995]{ken95} Kendall, T.R., Lennon, D.J., Brown, P.J.F., \& Dufton, P.L. 1995, A\&A, 298, 489
\bibitem[Kwok et al. (1997)]{kwo97} Kwok, S., Volk, K., \& Bidelman, W.P. 1997, ApJS, 112, 557
\bibitem[Lancon et al. (1992)]{lan92} Lancon, A., \& Rocca-Volmerange, B. 1992, A\&AS, 96, 593
\bibitem[Luhman \& Rieke (1998)]{luh98} Luhman, K.L., \& Rieke, G.H. 1998, ApJ, 497, 354
\bibitem[Koornneef 1983]{kor83} Koornneef, J. 1983, A\&A, 128, 84
\bibitem[McBreen et al. 1979]{mcb79} McBreen, B., Fazio, G.G., Stier, M., \& Wright, E.L. 1979, ApJ, 232, L183
\bibitem[Meyer et al. (1998)]{mey98} Meyer, M.R., Edwards, S., Hinkle, K.H., \& Strom, S.E. 1998, ApJ, 508, 397
\bibitem[Morgan \& Bally 1991]{mor91} Morgan, J.A., \& Bally, J. 1991, ApJ, 372, 505
\bibitem[Neckel 1978]{nec78} Neckel, T. 1978, A\&A, 69, 51
\bibitem[Osterloh et al. 1997]{ost97} Osterloh, M., Henning, Th., \& Launhardt, R. 1997, ApJS, 110, 71
\bibitem[Persi et al. 1996]{per96} Persi, P., Roth, M., Tapia, M., Marenzi, A.R., Felli, M., Testi, L., \& Ferrari-Toniolo, M. 1996, A\&A, 307, 591
\bibitem[Persson \& Campbell 1987]{per87} Persson, S.E., \& Campbell, B. 1987, AJ, 94, 416
\bibitem[Porter et al. 1998]{por98} Porter, J.M., Drew, J.E., \& Lumsden, S.L. 1998, A\&A, 332, 999
\bibitem[Shu et al. 1987]{Shu87} Shu, F.H., Adams, F.C., \& Lizano, S. 1987, ARA\&A, 25, 23
\bibitem[Straw et al. 1989]{str89} Straw, S.M., Hyland, A.R., \& McGregor, P.J. 1989, ApJS, 69, 99
\bibitem[Tapia et al. 1996]{tap96} Tapia, M., Persi, P., \& Roth, M. 1996, A\&A, 316, 102 
\bibitem[Walsh et al. (1997)]{wal97} Walsh, A.J., Hyland, A.R., Robinson, G., \& Burton, M.G. 1997, MNRAS, 291, 261
\bibitem[Walsh et al. 1999]{wal99} Walsh, A.J., Burton, M.G., Hyland, A.R., \& Robinson, G. 1999, MNRAS, 309, 905
\bibitem[Watson \& Hanson (1997)]{wat97} Watson, A.M., \& Hanson, M.M. 1997, ApJ, 490, L165
\bibitem[Wood \& Churchwell 1989]{woo89} Wood, D.O.S., \& Churchwell, E. 1989, ApJ, 340, 265

\end{thebibliography}
\end{document}